\documentclass[12pt,preprint]{aastex}
\def\apjs{{\rm ApJS}}                  % Astrophys. J. Supl.
\def\apjl{{\rm ApJ Let}}               % Astrophys. J. (Letters)
\def\Msol{\thinspace\hbox{$\hbox{M}_{\odot}$}}

\def\a4{\hsize 17.0cm \vsize 25.cm}
\newcommand{\der}[2]  { \frac{{\rm d}#1}{{\rm d}#2} }

\shorttitle{Radiation pressure and the Gas Dynamics Around YSCs}
\shortauthors{Silich \& Tenorio-Tagle}

\begin{document}

\title{How Significant is Radiation Pressure in the Dynamics of the Gas 
       Around Young Stellar Clusters?}

\author{Sergiy Silich}
\and
\author{Guillermo  Tenorio-Tagle} 
\affil{Instituto Nacional de Astrof\'\i sica Optica y
Electr\'onica, AP 51, 72000 Puebla, M\'exico; silich@inaoep.mx}

\begin{abstract}
The impact of radiation pressure on the dynamics of the gas in the vicinity 
of young stellar clusters is thoroughly discussed. The radiation over the 
thermal/ram pressure ratio time evolution is calculated explicitely 
and the crucial role of the cluster mechanical power and of the strong time
evolution of the ionizing photon flux and of the bolometric luminosity of the 
exciting cluster is stressed. It is shown that radiation has only a narrow 
window of opportunity to dominate the wind-driven shell dynamics. This may 
occur only at early stages of the bubble evolution and if the shell expands 
into a dusty and/or a very dense proto-cluster medium. The impact of radiation
pressure on the wind-driven shell becomes always negligible after about 3~Myr.
Finally, the wind-driven model results allow one to compare the model 
predictions with the distribution of thermal pressure derived from X-ray 
observations. The shape of the thermal pressure profile allows then to 
distinguish between the energy and the momentum dominated regimes of expansion
and thus conclude whether radiative losses of energy or the leakage of hot gas
from the bubble interior have been significant during the bubble evolution.
\end{abstract}

\keywords{galaxies: star clusters: general -- HII regions -- 
          hydrodynamics -- ISM: bubbles -- ISM: kinematics and dynamics}

\section{Introduction}

Several driving mechanisms have been proposed to explain the origin and
evolution of large expanding structures discovered in deep Halpha images 
and HI surveys of the Milky Way and other local group galaxies. The most 
thoroughly discussed model
involves the thermalization of the kinetic energy released by stellar winds 
and supernovae explosions inside a compact star cluster 
(\citealp{1987ApJ...317..190M, 1988ApJ...324..776M}). While this model is 
broadly consistent with observations of many bubbles in the visible and X-ray 
emission, several others driving forces, such as radiation pressure from the 
field stars (\citealp{1982ApJ...253..666E}) or the encounter of a high velocity 
cloud with the galactic disk (\citealp{1981A&A....94..338T}), were also 
discussed in the literature (see for a review \citealp{1988ARA&A..26..145T}).
The impact of radiation pressure on the dynamics of giant HII regions has been 
discussed recently by several authors who claimed that radiation pressure may 
play a crucial role in shaping the interstellar medium (ISM) around young 
massive clusters.  For example, \citet{2011ApJ...731...91L} compared the 
thermal pressure in the X-ray emitting plasma with the flux of radiation 
energy in the 30~Dor region and claimed that the hot gas pressure is generally 
weak and not dynamically important. \citet{2012ApJ...757..108Y} reached the 
same conclusion by making use the ionization parameter technique and claimed 
that in many individual and averaged over the galactic scale targets the 
shocked wind pressure cannot be large compared to radiation pressure. However 
\citet{2011ApJ...738...34P} measured the ionization parameter across the 30~Dor 
region and concluded that radiation pressure does not currently play a major 
role in star forming regions such as 30~Dor, although it may have been an 
important factor during the early evolution. The revision of the classical 
model is thus required for better understanding of such controversial results.

The development of HII regions begins right from the formation of the 
stellar cluster, when the ample supply of ionizing photons leads to a 
supersonic weak R ionization front (\citealp{1954BAN....12..187K, 
1976A&A....53..411T}), supersonic with respect to the neutral gas ahead and 
also supersonic with respect to the ionized gas behind it. The ionization 
front rushes then through the surrounding gas leaving it warm ($T \sim$ 
10$^4$ K) and ionized but dynamically unperturbed. This situation prevails  
throughout the formation phase, until the ionization front reaches the 
Str\"omgreen radius. The ionization front becomes then of D type with a strong 
shock progressing ahead of it into the surrounding gas. From then onwards and 
through the whole of the expansion phase, only a small fraction (less than 
1$\%$) of the ionizing photon flux will impinge on the ionization front and 
exert a pressure on the surrounding shell of swept up matter. For such a 
reason, in the classic HII region model one cannot use the full bolometric 
luminosity of the central cluster to estimate the strength of radiative 
pressure during the expansion of an HII region. \citet{2009ApJ...703.1352K}
assumed that radiation pressure causes the ionized gas to pile up into a thin 
outer shell and then concluded that while radiation pressure is generally not 
important for HII regions around a single or a handful of massive stars, it 
dominates the dynamics of giant HII regions driven by massive star clusters. 
\citet{2011ApJ...732..100D} calculated the density distribution in dusty HII 
regions with radiation pressure in the isothermal static approximation and 
showed that radiation pressure indeed produces a density gradient. However, 
the deviation from the homogeneous distribution is moderately small unless the 
parameter $\lambda_0 = Q_{49} n_{HII}$, where $Q_{49}$ is the number of 
ionizing photons in units of $10^{49}$~s$^{-1}$ and $n_{HII}$ is the mean 
density in the HII region, is large ($\lambda_0 \ge 10^4$~cm$^{-3}$).

A more realistic and also very different situation arises if one considers 
also the mechanical power of the exciting cluster and the strong evolution 
that the ionizing photon flux, the bolometric and the mechanical luminosities 
suffer after the first supernova explosion. Soon after the cluster wind 
interacts with the ionized surrounding gas, a strong shock wave begins to 
form a secondary wind-driven shell with the surrounding swept up matter. As 
the situation evolves, the pressure in the inner shell becomes larger than 
that of the free wind and thus a second, reverse shock is established in order
to equalize and maintain an even pressure between the shocked wind and the 
swept up ionized gas. The wind-driven shell sooner or latter cools down and 
begins to recombine, depending on the value of the background density. Such 
recombinations deplete photons from the outer HII region and cause the 
ionization front to supersonically recede towards the cluster to finally become
trapped within the inner wind-driven shell. The contribution of the 
star cluster UV radiation to the expansion of the original HII region 
becomes then insignificant.  From then onwards the wind-driven shock 
travels through a warm neutral or molecular gas whose sound speed is much 
smaller than that in the ionized interstellar medium. This permits the
wind-driven bubble to survive for a much longer time and leads to the
regime which \citet{2001PASP..113..677C} named the ``best case for wind'' 
model. \citet{2001PASP..113..677C} concluded that in this case the dynamical
evolution of the nebula is dictated by the stellar wind, but noted
that the ``best case for wind'' scenario requires the product 
$L_{912} n_{ISM}$, where $L_{912}$ is the luminosity of the ionizing source in
the Lyman continuum and $n_{ISM}$ is the density of the ambient medium, to be 
large. In the case of a single massive star, this requires a very 
high density ambient medium as the bolometric luminosities are rather small 
even in the case of the most massive O-stars. \citet{2012MNRAS.421.1283A}
took into consideration the effects of mass loading and thermal conduction and 
developed a detailed numerical model for the stellar wind bubble around 
$\Theta^{\prime}$Ori C, the main exiting star of the Orion nebula. The
calculations confirmed the ``best case for wind'' scenario and demonstrated 
that radiative heating makes the ionized part of the wind-driven shell thick. 
However, neither the ionizing source time evolution nor the radiation pressure
effects have been taken into consideration in these calculations.

Here we consider the impact that radiation pressure provides
on bubbles inflated by winds driven by massive star clusters. In such
a case, the ionizing ($L_{912}$) and the bolometric 
luminosity of a typical coeval cluster fall down rapidly after 
the first supernova explosion, whereas the mechanical luminosity of the 
cluster is maintained at a nearly constant level. One can notice this in 
Figure 1 which presents the results from the Starburst99 synthetic model 
(\citealp{1999ApJS..123....3L}) for a $10^6$\Msol \, coeval stellar cluster 
with stellar metallicity $Z = 0.4Z_{\odot}$ and standard Kroupa initial mass 
function. We also display on the right-hand panel of this figure the 
$L_{bol}$ over $L_{mech}$ and the $L_{912}$ over $L_{mech}$ ratios as these 
quantities are central to our further discussion and lead us to conclude that 
radiation pressure has a poor impact on the expansion of the wind-driven 
shell and thus on the global dynamics of giant HII regions.
%---------------------------------------------------------------
\begin{figure}[htbp]
\plottwo{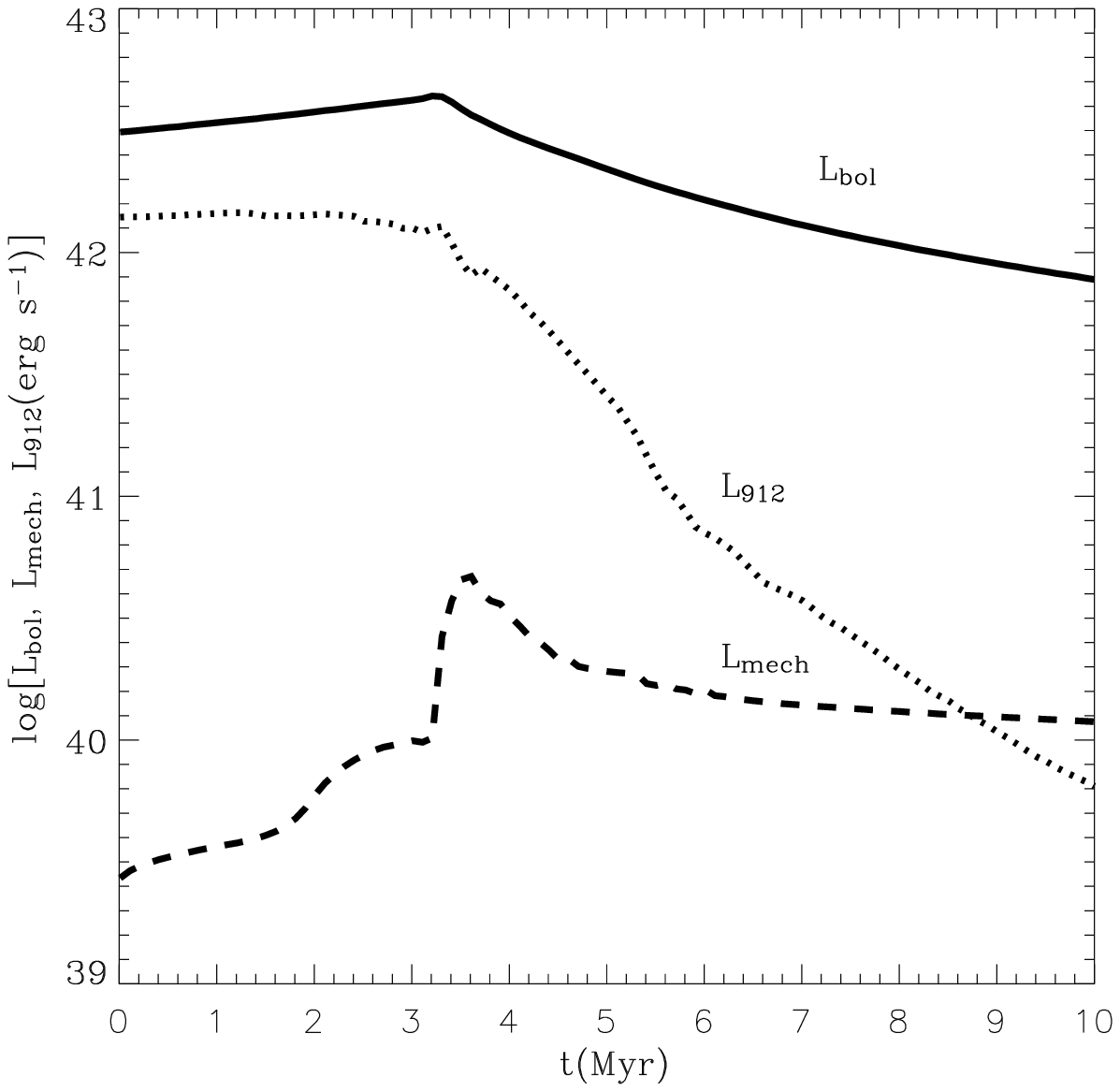}{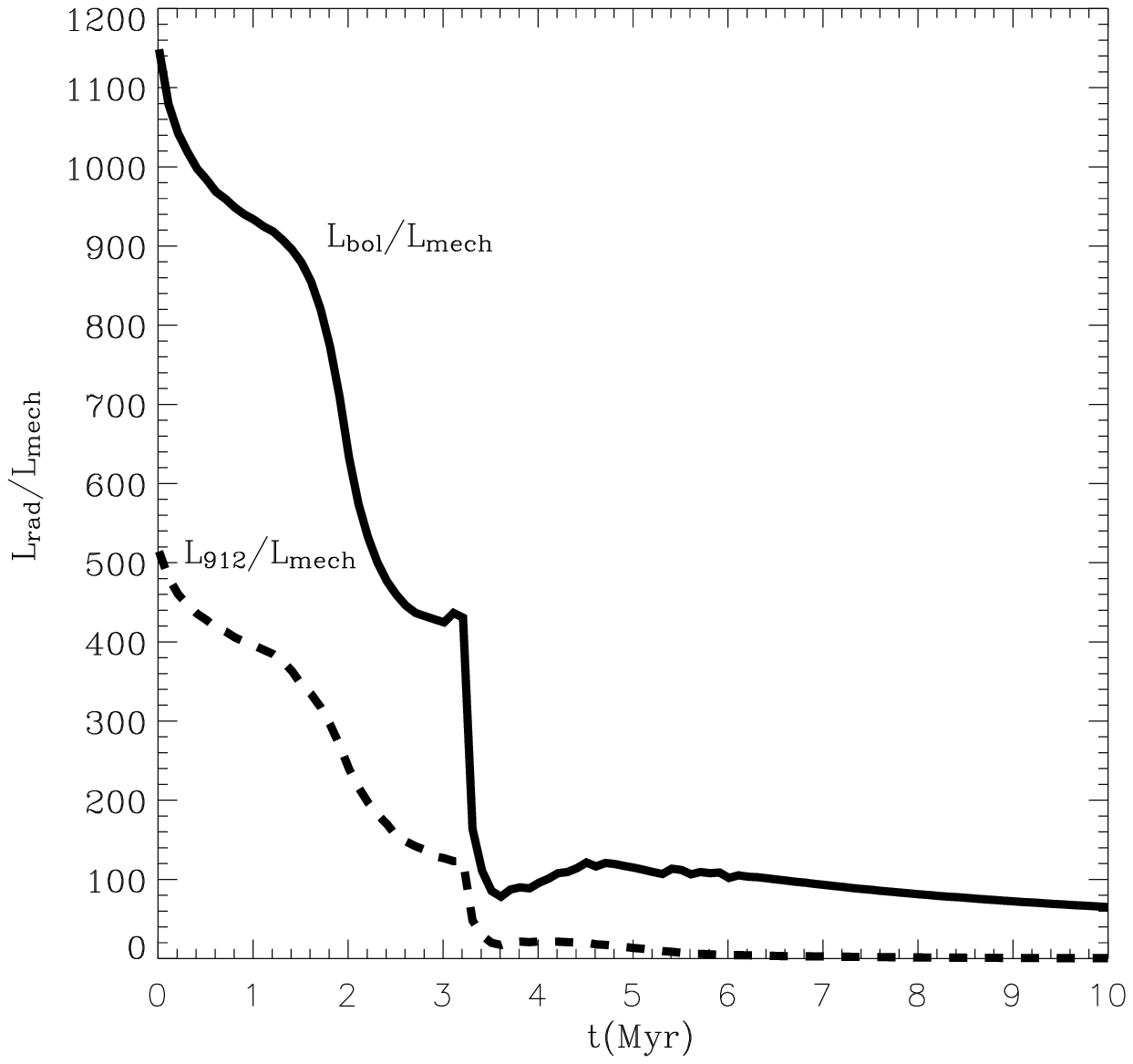}
\caption{The star cluster luminosities time evolution. The  bolometric,
Lyman continuum and  mechanical luminosities of a $10^6$\Msol \, coeval
stellar cluster with stellar metallicity $Z = 0.4Z_{\odot}$ and standard 
Kroupa initial mass function as predicted by the Starburst99, v6.0.3. The 
solid and dashed lines on the right-hand panel display the ratio of 
$L_{bol}$ over $L_{mech}$ and $L_{912}$ over $L_{mech}$ also plotted as a 
function of time.}
\label{fig1}
\end{figure}
%---------------------------------------------------------------   

The paper is organized as follows: the inner structure, the impact from 
radiative cooling and the time evolution of star cluster driven bubbles 
in the two possible (energy-dominated and momentum-dominated) hydrodynamic 
regimes are discussed in section 2. The impact of radiation pressure on the 
wind-driven shell is thoroughly discussed in section 3, where we calculate the 
radiation over the dynamical pressure ratio and discuss how this ratio evolves
with time. Section 4 presents the distribution of thermal pressure 
obtained from the wind-driven model and derives its appearance when projected
onto the plane of the sky. We also show in this section that the shape of the 
projected thermal pressure profile allows one to distinguish between bubbles 
evolving in the energy and in the momentum-dominated regimes. Our results are
compared with observations and other theoretical models in section 5 and
a summary of our major conclusions is given in section 6.

\section{The inner structure and evolution of star cluster driven bubbles}

The thermalization of the kinetic energy supplied by stellar winds and 
supernovae explosions within the exciting cluster leads to a large central 
pressure that causes the exit of the reinserted matter as a star cluster wind
(\citealp{1985Natur.317...44C,2000ApJ...536..896C,2007ApJ...658.1196T,
2011ApJ...743..120S}). Several physically distinct regions are then formed in 
the recently ionized gas (\citealp{1977ApJ...218..377W}). The innermost (free 
wind) zone is occupied by the high temperature reinserted plasma whose density, 
temperature and pressure asymptotically fall as $r^{-2}$, $r^{-4/3}$ and 
$r^{-10/3}$ while the wind velocity monotonically increases to reach its 
terminal value, $V_{\infty} \approx 3^{1/2} c_0$, where $c_0$ is the sound 
speed at the star cluster center (e.g. \citealp{2000ApJ...536..896C,
2011ApJ...743..120S}). The encounter of such high velocity 
outflow with the ambient, uniform density ionized medium, leads to the 
formation of the leading and reverse shocks. The leading shock sweeps up the 
ionized gas and compresses it into a secondary inner dense shell. The reverse 
shock decelerates and re-heats the free wind matter what results in a high 
thermal pressure in the shocked wind region. The thermal pressure in this 
zone is almost homogeneous as the temperature and the sound speed in the 
shocked wind plasma are large (\citealp{1977ApJ...218..377W}). Note that both, 
the free wind and the shocked wind regions are bright in the X-ray regime
(e.g. \citealp{1995ApJ...450..157C, 2000ApJ...536..896C, 2003MNRAS.339..280S,
2005ApJ...635.1116S}) due to the large temperature in these zones ($T \sim 
10^6$K - $10^7$K). Thus the interior of the secondary, wind-driven shell is 
usually transparent to the ionizing radiation from the central cluster.

\subsection{Energy dominated wind-driven bubbles}

The theory of interstellar wind-driven bubbles with a constant energy input 
rate was developed by  \citet{1977ApJ...218..377W}, \citet{1988ApJ...324..776M},
\citet{1992ApJ...388...93K} (see for a review \citealp{1995RvMP...67..661B}).
In this case the radius, $R_{b}$, the expansion velocity of the wind-driven
shell, $V_{b}$, and the thermal pressure inside the bubble volume, $P_{b}$, 
are:
%---------------------------------------------------------------
\begin{eqnarray}
      \label{eq2a}
      & & \hspace{-1.1cm} 
R_{b} = \left[\frac{375(\gamma-1) L_{mech}}
                      {28(9\gamma-4)\pi\rho_{ISM}}\right]^{1/5} t^{3/5} ,
      \\[0.2cm]     \label{eq2b}
      & & \hspace{-1.1cm}
V_{b} = \frac{3}{5} \frac{R_{b}}{t} ,
      \\[0.2cm]     \label{eq2c}
      & & \hspace{-1.1cm}
P_{b} = 7 \rho^{1/3}_{ISM} \left[\frac{3(\gamma-1) L_{mech}}
                  {28(9\gamma-4) \pi R^2_{b}}\right]^{2/3} ,
\end{eqnarray}
%---------------------------------------------------------------
where $t$ is the dynamical time, $\rho_{ISM}$ is the ambient gas density and 
$L_{mech}$ is the mechanical luminosity of the central cluster. 
One can find the location of the reverse shock from the relation $P_b = 
P_{ram}$, where $P_{ram} = \rho_w V^2_{\infty}$ is the free wind 
ram pressure and the density in the wind, $\rho_w$, is calculated at 
the reverse shock position: $\rho_w = L_{mech} / 2 \pi R_{RS}^2 V^3_{\infty}$ 
(e.g. \citealp{1977ApJ...218..377W}):
%---------------------------------------------------------------
\begin{equation}
      \label{eq5}
R_{RS} = \left(\frac{L_{mech}}{2 \pi V_{\infty} P_b}\right)^{1/2} .
\end{equation}
%---------------------------------------------------------------

The wind-driven shell of swept up interstellar matter is initially adiabatic 
and hot and thus transparent to the ionizing radiation. However, it cools 
down in a short time scale (\citealp{1988ApJ...324..776M}):
%---------------------------------------------------------------
\begin{equation}
      \label{eq6}
\tau_1 = (2.3 \times 10^4) Z^{-0.42}_{ISM} n^{-0.71}_{ISM} 
              L^{0.29}_{38} yr,
\end{equation}
%---------------------------------------------------------------
where $n_{ISM}$ and $Z_{ISM}$ are the atomic number density and metallicity
in the surrounding medium, $L_{38}$ is the star cluster mechanical luminosity 
in units of $10^{38}$~erg s$^{-1}$ and $\gamma = 5/3$. As soon as the 
wind-driven shell cools down, it recombines and begins to absorb ionizing 
photons emerging from the central cluster. This reduces the number of Lyman 
continuum photons in the outer HII region. From then onwards, the ionization 
front detaches from the outer shell and moves back towards the wind-driven 
shell reaching it when this grows thick enough as to absorb all ionizing 
photons from the central cluster. One can obtain the characteristic trapping 
time, $\tau_{trap}$, from the condition (\citealp{1997A&A...326.1195C}):
%---------------------------------------------------------------
\begin{equation}
      \label{eq8}
N_{912} = 4 \pi R_{b}^2 \Delta R \beta n^2_s .
\end{equation}
%---------------------------------------------------------------
where $N_{912}$ is the number of Lyman continuum photons, $\beta = 2.59 
\times 10^{-13}$~cm$^3$ s$^{-1}$ is the recombination coefficient to all but 
the ground level. We assume that thermal pressure in the ionized shell 
is uniform and equal to that in the shocked wind region (see Arthur, 2012). 
The density of the ionized shell then is $n_s = P_b / \mu_i c^2_{HII}$, 
where $\mu_i = 14 m_H / 11$ is the mean mass per ion, $c_{HII} = 
(k T_{HII} / \mu_t)^{1/2}$ is the isothermal speed of sound in the ionized 
gas and $\mu_t = 14/23 m_H$ is the mean mass per particle in the completely 
ionized plasma with one helium atom per every ten hydrogen atoms. 
The shell thickness, $\Delta R$, is calculated from mass conservation: 
%---------------------------------------------------------------
\begin{equation}
      \label{eq8a}
\Delta R = \frac{n_{ISM}}{n_s} \frac{R_b}{3} .
\end{equation}
%---------------------------------------------------------------
Equations (\ref{eq2a}), (\ref{eq2b}), (\ref{eq8}) and (\ref{eq8a}) then 
yield:
%---------------------------------------------------------------
\begin{equation}
      \label{eq9}
\tau_{trap} = \frac{(9\gamma-4)}{5(\gamma-1)} \frac{\mu_i c^2_{HII} N_{912}}
                {\beta n_{ISM} L_{mech}} .
\end{equation}
%---------------------------------------------------------------
During the forthcoming evolution the HII region is bounded by the wind-driven 
shell. The leading shock then remains strong even when the shell velocity 
drops below $c_{HII}$  as the leading shock Mach number must be calculated 
with respect to the sound speed in the neutral ambient medium.
Note, that the ionizing radiation may be trapped within the shell at much 
earlier times if the shell is dusty (see \citealp{2011ApJ...735...66M}).

Similarly, one can find the relative thickness of the ionized shell
$\Delta R / R_b$ before and after the trapping time $\tau_{trap}$. For 
$t \le \tau_{trap}$ the shell is completely photoionized and the value of 
$\Delta R / R_b$ is restricted by the swept up mass $M_{shell} = 
4 \pi R^3_b \rho_{ISM} /3$ while for $t > \tau_{trap}$ the shell is partially
ionized and the thickness of the ionized layer depends on the number of 
incident ionizing photons $N_{912}$:
%---------------------------------------------------------------
\begin{eqnarray}
      \label{eq9a}
      & & \hspace{-1.1cm} 
\Delta R/ R_{b} = \frac{\mu_i c^2_{HII} n_{ISM}}{3 P_b} \qquad 
                   t \le \tau_{trap} ,
      \\[0.2cm]     \label{eq9b}
      & & \hspace{-1.1cm}
\Delta R/ R_{b} = \frac{\mu^2_i c^4_{HII} N_{912}}{4 \pi P^2_b R^3_b} 
                  \qquad t > \tau_{trap} .
\end{eqnarray}
%---------------------------------------------------------------  

Figure 2 shows the wind-driven shell evolution onto an ambient medium
with densities $n_{ISM} = 1$~cm$^{-3}$ and $n_{ISM} = 1000$~cm$^{-3}$ (solid 
and dashed lines, respectively). The selected mechanical luminosity of the 
driving cluster is $L_{mech} = 10^{40}$~erg s$^{-1}$ in both cases, which is 
the average value for a young $10^6$\Msol \, cluster with a standard Kroupa 
initial mass function and 0.4Z$_{\odot}$ metallicity 
(\citealp{1999ApJS..123....3L}) whereas the number of ionizing photons changes 
with time as is predicted by the evolutionary synthetic models (see Figure 1).
It will be shown below that this approximation does not affect our major
results significantly because the radiative over dynamical pressure ratio
depends mainly on the bolometric over mechanical luminosity ratio and
only weakly on the star cluster mechanical luminosity itself (see section 3).
In the lower density case the wind-driven shell cools after $\tau_1 
\approx 10^5$~yr and grows thick enough to absorb all ionizing photons 
($\sim 10^{53}$~s$^{-1}$) at about $\tau_{trap} \approx 3.4$~Myr 
(see Figure 2). At that time the radius of the shell reaches about 360~pc and 
its expansion velocity has dropped to about 60~km s$^{-1}$. Also, the number 
of Lyman continuum photons emerging from the central cluster begins to 
decrease due to the explosion of the most massive stars (see Figure 1). 
%---------------------------------------------------------------
\begin{figure}[htbp]
\plottwo{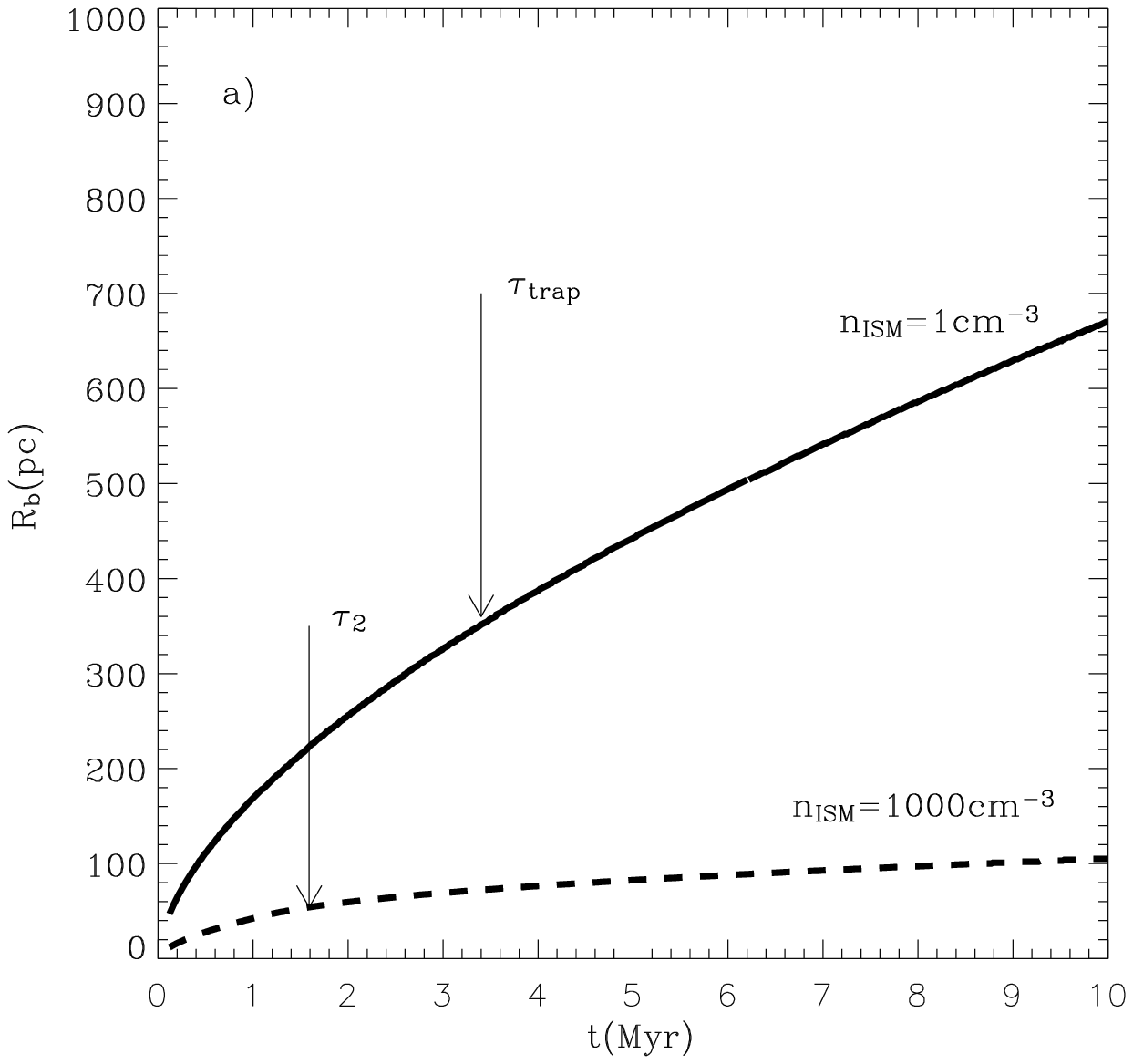}{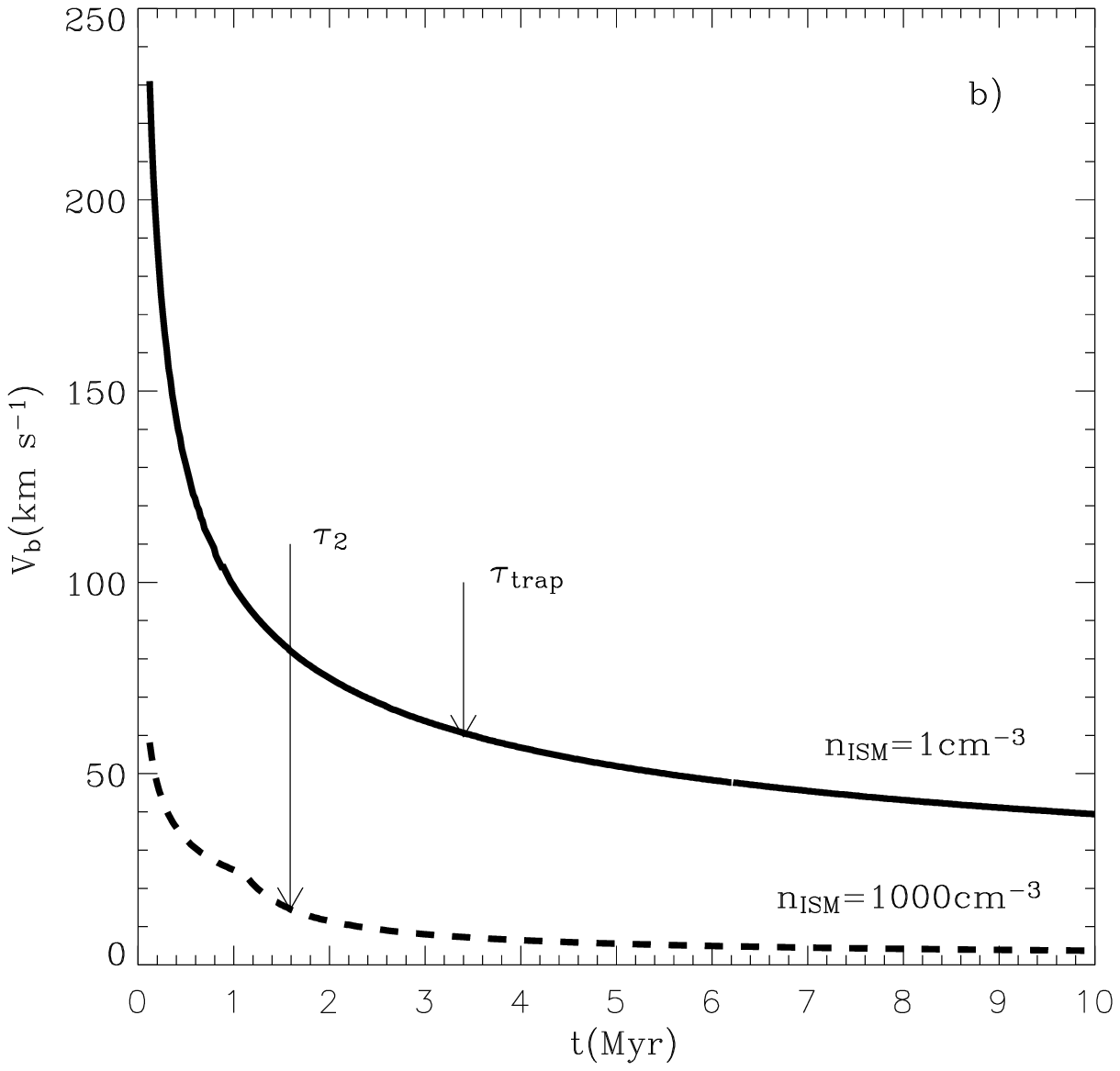}
\caption{The wind-driven bubble evolution. Panels a and b present the radius 
and the velocity of the shell when it expands into an ambient medium
with densities: $n_{ISM} = 1$~cm$^{-3}$ and $n_{ISM} = 1000$~cm$^{-3}$ (solid 
and dashed lines) respectively. The mechanical luminosity of the driving 
cluster is the same in both cases: $L_{mech} = 10^{40}$~erg s$^{-1}$. The 
metallicity of the ambient gas is $Z_{ISM} = 0.4Z_{\odot}$, the rate of
ionizzing photons was calculated by making use of Starburst99 synthesis 
model. The arrows mark the characteristic trapping time, $\tau_{trap}$, in 
the lower density case and the characteristic time scale for the shocked wind 
gas to cool down, $\tau_2$, (see sections 2.1 and 2.2 for more details) in 
the larger density case, respectively.} 
\label{fig2}
\end{figure}
%--------------------------------------------------------------- 

\subsection{Momentum dominated wind-driven bubbles}

It is likely that young stellar clusters are embedded initially into dense 
parental clouds. Therefore Figure 2 also presents the wind-driven bubble 
evolution into a dense ambient medium with $n_{ISM} = 1000$~cm$^{-3}$ (dashed 
lines). In this case the shell absorbs all ionizing photons as soon as it 
grows enough to encompass all the cluster. Thus, in the high density 
environment the shell cools down and traps the ionizing radiation from the 
driving cluster very rapidly, much faster than when it expands into a typical 
ISM with $n_{ISM} = 1$~cm$^{-3}$. Besides this, in the denser environment the 
shocked wind region may lose a significant fraction of the deposited energy 
in a short time scale due to strong radiative cooling in the conduction
zone dominated by mass evaporated from the wind-driven shell 
(\citealp{1988ApJ...324..776M}):
%---------------------------------------------------------------
\begin{equation}
      \label{eq15}
\tau_2 = (1.6 \times 10^7) Z^{-35/22}_{ISM} n^{-8/11}_{ISM} 
              L^{3/11}_{38} yr .
\end{equation}
%---------------------------------------------------------------
In the case under consideration this occurs at $\tau_2 \approx 1.6$~Myr, when 
the shell radius reaches about 60~pc and the expansion velocity is $\sim 
20$~km s$^{-1}$. Without a pressure support, the reverse shock moves rapidly 
towards the shell and the further expansion is then due to the direct impact 
of the free wind momentum. This may also be the case if a collisionless 
reverse shock at $R_{RS} < R_b$ is not formed (see the discussion of the 
reverse shock conditions in \citealp{2001PASP..113..677C}). One can neglect 
this transition phase and compute the shell dynamics in the momentum dominated
regime by writing down the momentum equation, assuming that the transition to 
the momentum dominated regime occurs at  $t = \tau_2$:
%---------------------------------------------------------------
\begin{equation}
      \label{eq12}
M_{sh}(t) \der{R_b}{t} =  M_{sh}(\tau_2) V_b(\tau_2) +
          \int_{\tau_2}^t {\dot M_w} V_{\infty} {\rm d} t^{\prime} ,  
\end{equation}
%---------------------------------------------------------------
where $M_{sh}(t) = 4 \pi R^3_b(t) \rho_{ISM} / 3$ is the mass of the shell at 
time $t$ and ${\dot M_w} = 2 L_{mech} / V^2_{\infty}$ is the rate of mass 
deposition due to the wind.

The integration of equation (\ref{eq12}) yields:
%---------------------------------------------------------------
\begin{eqnarray}
      \label{eq13a}
      & & \hspace{-1.1cm} 
R_{b} = R_{c}
\left[\frac{3 L_{mech} (t^2 + \tau^2_2)}
      {\pi V_{\infty} \rho_{ISM} R^4_{c}} +
      \left(\frac{12}{5} - \frac{6 L_{mech} \tau^2_2}
      {\pi V_{\infty} \rho_{ISM} R^4_{c}}\right) \frac{t}{\tau_2} -
      \frac{7}{5}\right]^{1/4} \, ,
      \\[0.2cm]     \label{eq13b}
      & & \hspace{-1.1cm}
V_{b} = \frac{3 L_{mech}}{2 \pi V_{\infty} \rho_{ISM} R^3_b} t +
        \left(\frac{3}{5} \frac{R^4_{c}}{\tau_2} -
        \frac{3}{2} \frac{L_{mech} \tau_2}{\pi V_{\infty} \rho_{ISM}}
        \right) R^{-3}_b        
      \\[0.2cm]     \label{eq13c}
      & & \hspace{-1.1cm}
P_{ram} = \frac{L_{mech}}{2 \pi V_{\infty} R^2_b} ,
\end{eqnarray}
%---------------------------------------------------------------
where $R_{c}$ is the radius of the shell at the time when the 
transition  to the momentum dominated regime occurs: $R_c = R_b(\tau_2)$
and $R_b(\tau_2)$ is calculated by means of equation (\ref{eq2a}). Note, that
if $R_c \to 0$ and $\tau_2 \to 0$, equation (\ref{eq13a}) is reduced to a well
known relation $R_b \sim t^{1/2}$ (e.g. equation 3.1 in 
\citealp{1992ApJ...388...93K}).

Note that in the momentum-dominated regime the value of the thermal pressure 
in the wind-driven shell is equal to the wind ram pressure at the shell inner
edge. The relative thickness of the ionized layer then is:
%---------------------------------------------------------------
\begin{equation}
      \label{eq12a}
\Delta R / R_b =  \frac{\pi \mu^2_i V^2_{\infty} c^4_{HII} R_b N_{912}}
                  {\beta L^2_{mech}} .
\end{equation}
%---------------------------------------------------------------

If thermal conduction and mass evaporation from the cold shell are inhibited 
by magnetic fields (\citealp{1962pfig.book.....S}), the radiative losses of 
energy from the shocked wind region are much smaller. In this case the ion 
number density and temperature in the shocked wind region are:
%---------------------------------------------------------------
\begin{eqnarray}
      \label{eq16a}
      & & \hspace{-1.1cm} 
n_b = \frac{3 {\dot M} t}{4 \pi \mu_i R^3_b} = \frac{3 L_{mech} t}
           {2 \pi \mu_i V^2_{\infty} R^3_b} ,
      \\[0.2cm]     \label{eq16b}
      & & \hspace{-1.1cm}
T_b =  \frac{\mu_t}{\mu_i} \frac{P_b}{k n_b} = 
       \frac{5(\gamma-1)\mu_t V^2_{\infty}}{2(9\gamma-4) k} ,
\end{eqnarray}
%---------------------------------------------------------------
Thus temperature in the shocked wind region in this case depends only on the 
wind terminal speed and does not change with time. If $V_{\infty} = 
1000$~km s$^{-1}$, $T_b \approx 1.1 \times 10^7$~K.

Radiative cooling begins to reduce the thermal pressure in the shocked wind 
zone and affect the bubble dynamics only when radiative losses of energy from 
the bubble interior $Q = 4 \pi n^2_b \Lambda(T_b, Z_{b}) R^3_b / 3$ 
exceed the energy input rate from the central cluster: $Q > L_{mech}$. This 
occurs if the evolutionary time $t$ grows larger than
%---------------------------------------------------------------
\begin{equation}
      \label{eq17}
\tau_3 = \left(\frac{\mu^2_i V^4_{\infty}}
                      {3 \Lambda(T_b,Z_b)}\right)^5
            \left(\frac{\pi}{L_{mech}}\right)^2
           \left[\frac{375(\gamma-1)}{28(9\gamma-4)\rho_{ISM}}\right]^3 ,
\end{equation}
%--------------------------------------------------------------- 
where $Z_b$ is the gas metallicity in the shocked wind region and 
$\Lambda(T_b,Z_b)$ is the cooling function. One can obtain from equation
(\ref{eq17}) that $\tau_3$ is much longer than the characteristic life-time
of giant HII regions ($\sim 10$~Myr). This implies that the radiative losses 
of energy from the shocked wind region in this case are insignificant and 
the wind-driven bubble expands in the energy dominated regime during the 
whole evolution of the HII region.

The reverse shock may also be closer to the  wind-driven shell than one would 
expect from equation (\ref{eq5}), if the hot, shocked wind plasma escapes 
from the bubble into the surrounding low density medium, as suggested in 
\citet{2011ApJ...731...91L}. In this case the expansion velocity of the 
wind-driven shell differs from that predicted by the energy dominated model 
(equations \ref{eq2a} - \ref{eq2c}) and may be more similar to what the 
momentum dominated model predicts.

\section{The impact of radiation pressure on the wind-driven shell}

The radiation pressure on the wind-driven shell is:
%---------------------------------------------------------------
\begin{equation}
      \label{eq7}
P_{rad} = f_{trap} L_{bol} / 4 \pi c R^2_{b} ,
\end{equation}
%---------------------------------------------------------------
where $f_{trap}$ is the fraction of the bolometric luminosity absorbed within 
the shell, $c$ is the speed of light and $R_{b}$ is the radius of the shell. 
In the energy dominated regime equations (\ref{eq5}) and (\ref{eq7}) yield: 
%---------------------------------------------------------------
\begin{equation}
      \label{eq10}
\epsilon = \frac{P_{rad}}{P_{b}} = \frac{f_{trap}}{2} 
           \left(\frac{L_{bol}}{L_{mech}}\right) 
           \left(\frac{V_{\infty}}{c}\right)
           \left(\frac{R_{RS}}{R_{b}}\right)^2 .
\end{equation}
%---------------------------------------------------------------
One can note from relation (\ref{eq10}) that the small $V_{\infty}$ over $c$ 
ratio and the fact that $R_{RS}$ is usually much smaller than $R_b$, decrease 
significantly $\epsilon$ from what one would expect during the early bubble 
evolution when the bolometric luminosity of the driving cluster exceeds the 
mechanical energy input rate by a factor of $\sim 1000$ (see Figure 1).
Hereafter we assume in our calculations that $V_{\infty} = 1000$~km 
s$^{-1}$.

The $P_{rad}$ over $P_b$ ratio may be presented in a different form which
shows explicitely how it evolves with time, if one makes use of equation 
(\ref{eq2c}) instead of equation (\ref{eq5}):
%---------------------------------------------------------------
\begin{equation}
      \label{eq11}
\epsilon = \frac{P_{rad}}{P_{b}} = \frac{f_{trap}}{28 \pi c}
           \left[\frac{28(9\gamma-4) \pi}{3(\gamma-1)}\right]^{4/5}
           \left(\frac{L_{mech}}{25 \rho_{ISM}}\right)^{1/5}
           \frac{L_{bol}}{L_{mech}} \, t^{-2/5} .
\end{equation}
%---------------------------------------------------------------
Equation (\ref{eq11}) shows that the ratio of the bolometric over mechanical 
luminosity is the major factor which decides which of the two driving forces, 
radiation or thermal pressure, dominates the shell dynamics and that the 
$P_{rad}$ over $P_{b}$ ratio has only a weak dependence on the ambient gas 
density and on the star cluster mechanical luminosity. 

The relation between $P_{rad}$ and $P_{b}$ will be different if the shell 
expands in the momentum dominated regime. In this case, one has to compare 
the radiation and the free wind ram pressure at the inner edge of the shell: 
%---------------------------------------------------------------
\begin{equation}
      \label{eq14}
\epsilon = \frac{P_{rad}}{P_{ram}} = 
           \frac{f_{trap}}{2} \frac{V_{\infty}}{c} \frac{L_{bol}}{L_{mech}} . 
\end{equation}
%--------------------------------------------------------------- 
Note, that relationship (\ref{eq14}) is universal as it does not depend 
on the ambient gas density and the $L_{bol}$ over $L_{mech}$ ratio is a 
universal function of time for all clusters with a given initial mass 
function and metallicity.   

Figure 3 presents the $P_{rad}$ over $P_b$ ratio time evolution for the two 
cases here considered. As mentioned above, it was assumed that the 
$L_{bol}$ over $L_{mech}$ changes with time as predicted by the evolutionary 
synthesis model Starburst99 but a constant value of the mechanical luminosity,
$L_{mech} = 10^{40}$~erg s$^{-1}$, was used in the calculations as $\epsilon$ 
depends on the value of $L_{mech}$ itself only in the energy dominated regime 
and even then this dependence is weak (see equations \ref{eq11} and 
\ref{eq14}). Numerical solutions with Starburst99-generated input mechanical 
luminosity were used by \citet{2005ApJ...619..755D} in their discussion of the 
synthetic spectral energy distribution in starbursts. 

Note that $P_b$ in Figure 3 is the thermal pressure 
in the shocked wind region in the energy dominated case and $P_b = P_{ram}$ in 
the momentum dominated regime, respectively. In the low density case (solid 
line) the  $P_{rad}$ over $P_b$ ratio is larger than unity only at 
very early times ($t \le 2 \times 10^5$~yr) and only if the shell is dusty 
and optically thick to the bolometric luminosity of the central cluster 
($f_{trap} = 1$). Otherwise it should grow thick enough to absorb all 
ionizing photons only after $\tau_{trap} = 3.4$~Myr, when the contribution
of radiation pressure to the shell dynamics is already negligible. 
The $P_{rad}$ over $P_b$ ratio is small during the initial shell 
evolution if the driving cluster is embedded into a high density medium (see 
dashed line in Figure 3 and equation \ref{eq11}). It then suddenly grows to 
a value of 1.4 (see equation \ref{eq14}) as the gas in the bubble interior 
cools down, the reverse shock reaches the shell and the further bubble 
evolution proceeds in the momentum dominated regime. 
However, even in this case the $P_{rad}$ over $P_b$ ratio 
becomes smaller than one after $\sim 2$~Myrs and thus radiation has only a 
narrow window of opportunity to dominate the shell dynamics.
Note also that the true value of $f_{trap}$ is set by the SED of the incident 
spectrum and by the optical depth of the shell and may be smaller than
unity. In this respect Figure 3 presents an upper limit for the $P_{rad}$ 
over $P_b$ ratio. For example, if the shell is thick to the Lyman 
continuum only, the $P_{rad}$ over $P_b$ ratio would be smaller than what 
is shown in Figure 3 within a factor of 2 - 4, depending on the star cluster 
age (compare the bolometric and the Lyman continuum luminosities in Figure 1). 
On the other hand, the adiabatic wind terminal speed $V_{\infty}$ also changes
with time and may be different within a factor of 3 from the selected value 
(see \citealp{2011ApJ...740.75W}).
%---------------------------------------------------------------
\begin{figure}[htbp]
\plotone{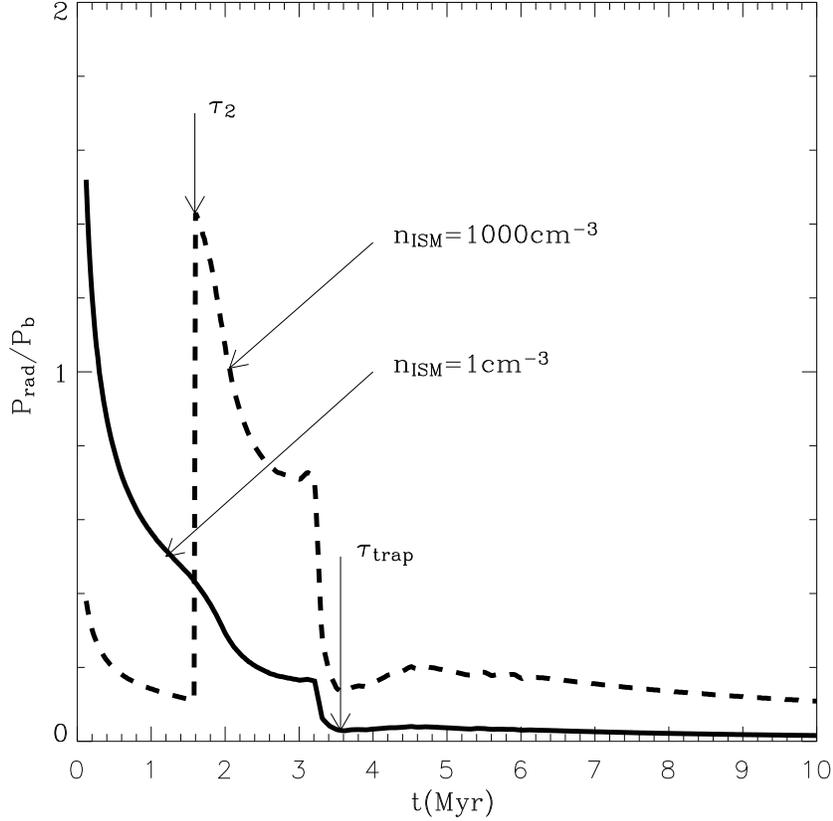}
\caption{The contribution of radiation pressure to the star cluster wind-driven
shell dynamics. Vertical arrows mark the time when the transition to the
momentum dominated regime occurs ($\tau_2$) and the trapping time 
for the ionizing radiation ($\tau_{trap}$) in the high and low density models,
respectively. Note that in the low density case (solid line) the wind-driven 
shell does not even trap all ionizing radiation during the early evolution 
($\tau_{trap} \approx 3.4$~Myr) if it is not dusty. The dashed line shows the 
$P_{rad}$ over $P_b$ ratio in the case when $n_{ISM} = 1000$~cm$^{-3}$.}
\label{fig3}
\end{figure}
%--------------------------------------------------------------- 

\section{Thermal pressure as an indicator of the hydrodynamic regime}

Here we calculate the distribution of thermal pressure inside a wind-driven 
bubble volume and compare it to its projection onto the plane of the sky as
it is done in high resolution studies of X-ray emission (e.g. 
\citealp{2011ApJ...731...91L}).

In order to calculate the distributions of physical quantities (density, 
temperature, thermal pressure, velocity) in the free wind region and localize 
the reverse shock position, we make use of \citet{2011ApJ...743..120S} star 
cluster wind driven model, which assumes that stars are exponentially 
distributed within the star cluster volume. The model input parameters are: 
the starburst mechanical luminosity, $L_{mech} = 10^{40}$~erg s$^{-1}$, the 
characteristic scale length of the stellar density distribution, $R_{core} = 
1$~pc, the adiabatic wind terminal speed, $V_{A\infty} =1000$~km s$^{-1}$, and 
the metallicity in the free wind region, which we set to be $Z_X = 
0.4 Z_{\odot}$. The position of the reverse shock at different times was 
calculated by means of equation (\ref{eq5}) which relates the thermal pressure 
in the shocked wind zone and the ram pressure in the free wind region. We then 
compute the density and the temperature in the shocked wind region by means of 
equations (\ref{eq16a}) and (\ref{eq16b}) and calculate the distribution of 
thermal pressure in the same way, as in \citet{2011ApJ...731...91L}.
Specifically, we calculate the emission measure and the waited temperature of 
the hot X-ray plasma along lines of sight with different impact parameters $X$ 
by integrating the model predicted density and temperature distributions. The 
emission measure and the waited temperature then are:
%----------------------------------------------------------------------
\begin{eqnarray}
      \label{eq15a}
      & & \hspace{-1.1cm} 
EM(X) = 2 \int_0^{L_{max}} n^2(r) {\rm d} l ,
      \\[0.2cm]     \label{eq15b}
      & & \hspace{-1.1cm}
 T(X) = 2 \, EM^{-1}(X) \int_0^{L_{max}} T(r) n^2(r) {\rm d} l \,  ,
\end{eqnarray}
%----------------------------------------------------------------------
where $l$ and $L_{max} = (R^2_{b} - X^2)^{1/2}$ are the distance and the path 
length along the line of sight, $n(r)$ and $T(r)$ are calculated at 
$r = (X^2 + l^2)^{1/2}$. One can obtain then the plasma density and the 
thermal pressure along lines of sight with an impact parameter $X$ from
the relations:
%---------------------------------------------------------------
\begin{equation}
      \label{eq10a}
n(X) = [EM(X) / 2 L_{max}]^{1/2} ,  \qquad
P(X) = \mu_i k n(X) T(X) / \mu_t .
\end{equation}
%---------------------------------------------------------------
Note, that the procedure takes into account that lines of sight cross the 
free wind region if the impact parameter $X$ is smaller than the reverse 
shock radius $R_{RS}$. We also take care to omit segments along lines of 
sight where the temperature drops below the X-ray cut-off temperature, 
$T_{cut} = 5 \times 10^5$~K (see Figure 4).

In the momentum-dominated regime the shocked wind region collapses and the star
cluster wind impacts directly on the shell. Strong radiative cooling reduces 
then the temperature immediately behind the reverse shock  leaving the shocked
wind matter photoionized and at $10^4$K. Thus, in the momentum-dominated
regime the contribution of the shocked wind gas to the X-ray emission 
is negligible.  

The distributions of thermal pressure obtained from the wind-driven bubble 
model and those, obtained  by integration along different lines of sight 
are compared in Figure 4 (upper and bottom panels, respectively). The middle 
panels present the temperature distributions, which one can obtain from the 
wind-driven bubble model. Panels on the left-hand and the right-hand sides 
compare profiles obtained in the low ($n_{ISM} = 1$~cm$^{-3}$) and high 
($n_{ISM} = 1000$~cm$^{-3}$) density cases. Solid and dashed lines show 
the distributions of thermal pressure and temperature at different
evolutionary times: $t = 2$~Myr and $t = 5$~Myr, respectively.
%---------------------------------------------------------------
\begin{figure}[htbp]
\vspace{18.5cm}
\includegraphics{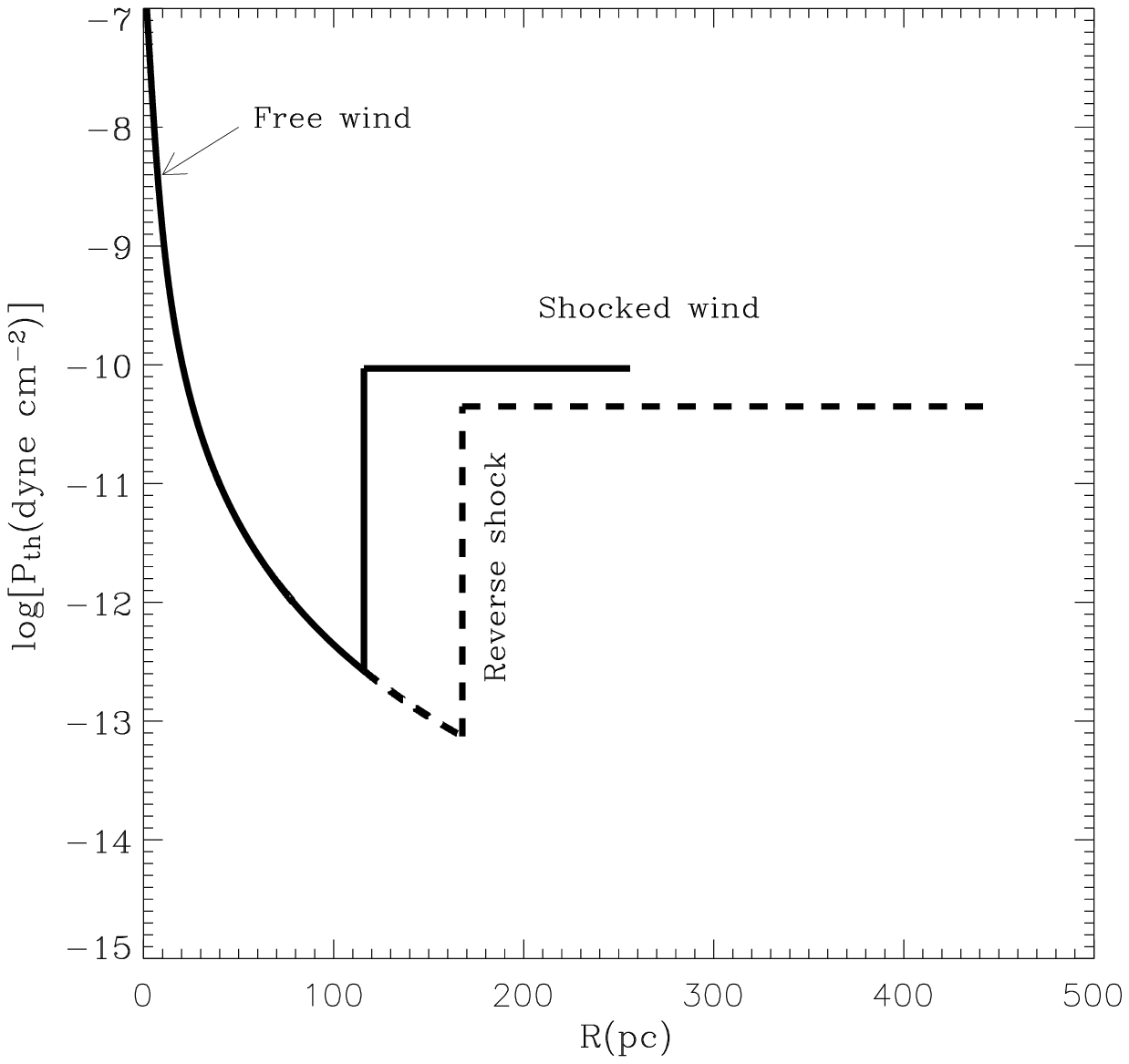}
\includegraphics{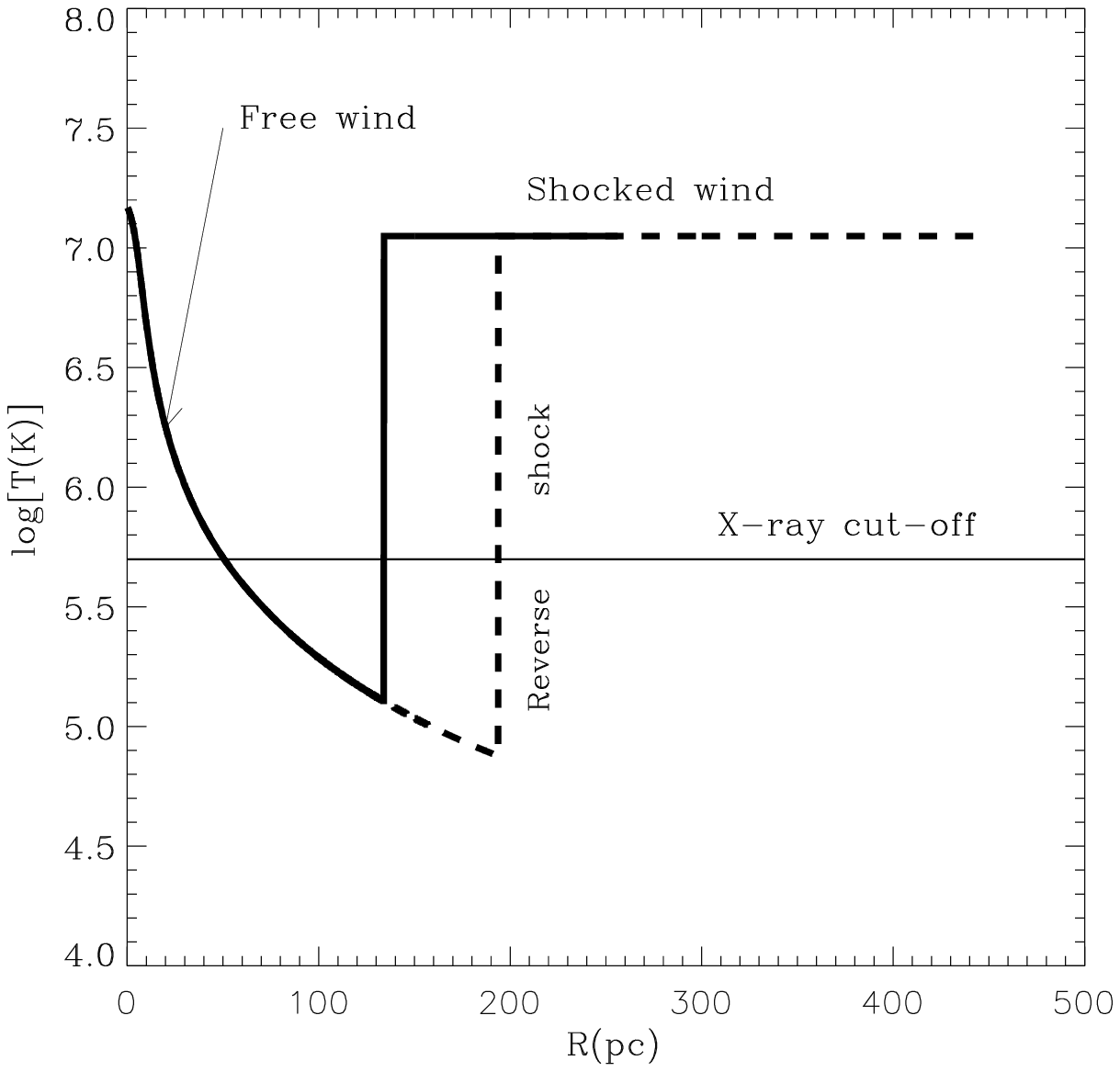}
\includegraphics{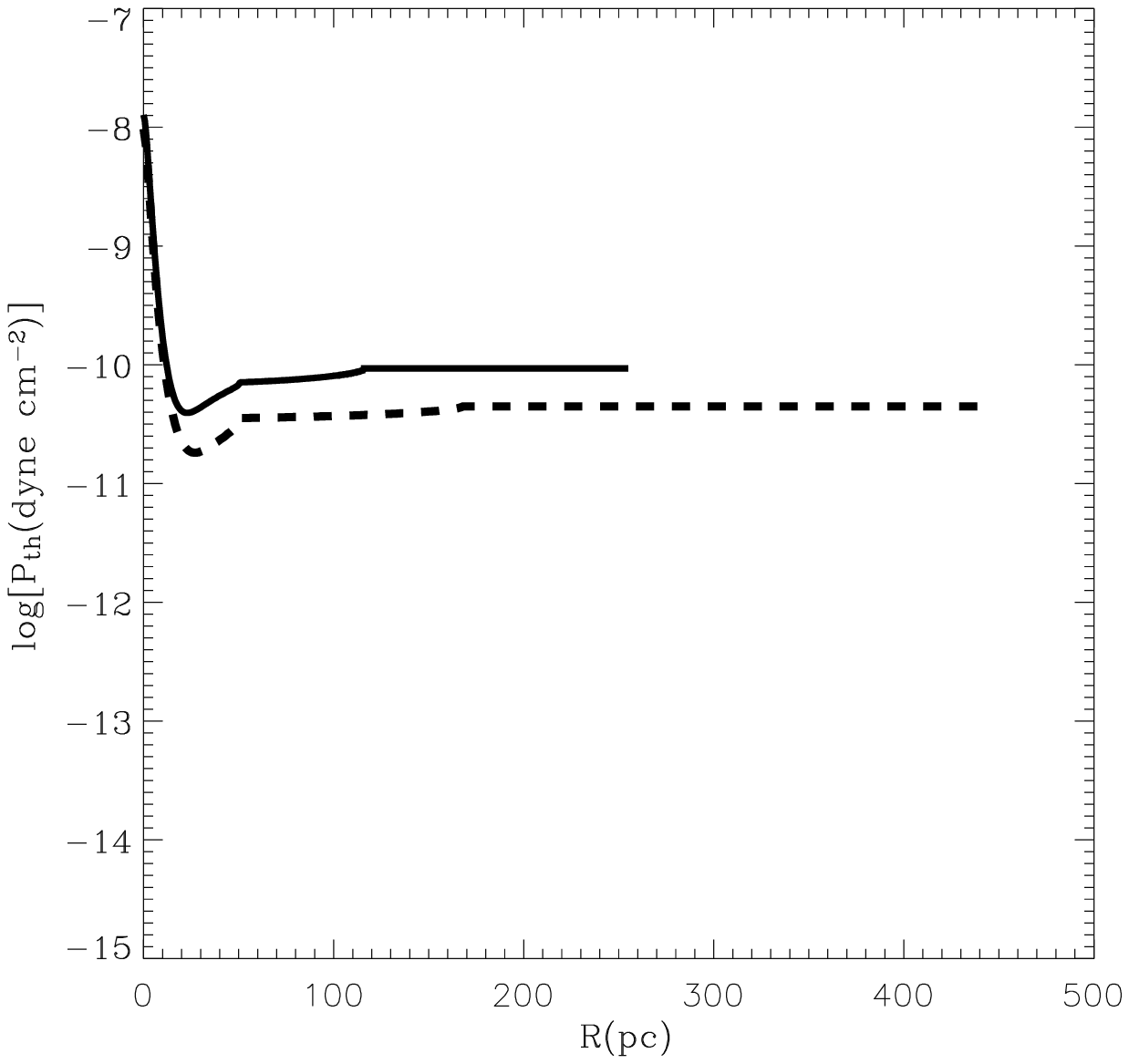}
\includegraphics{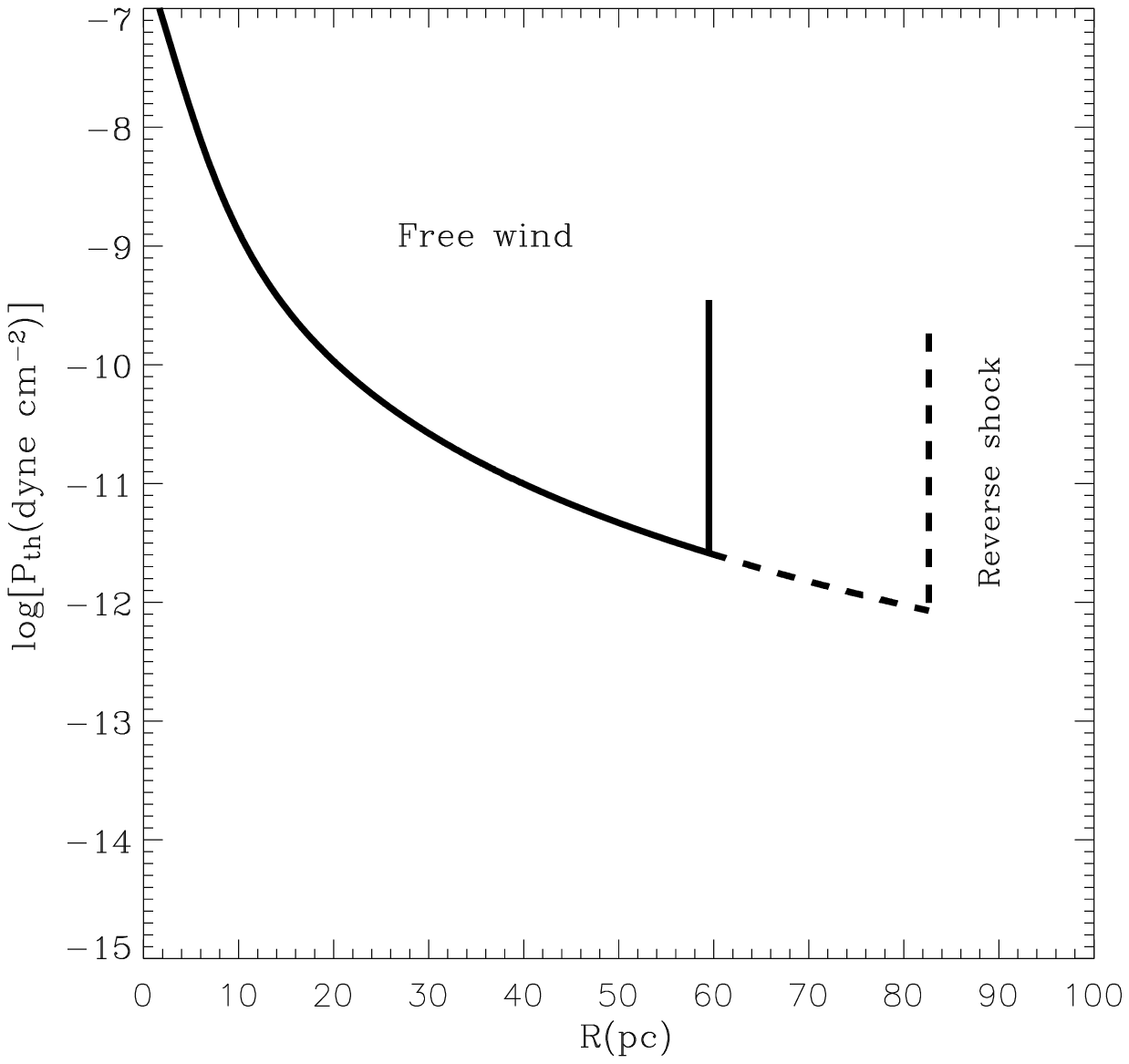}
\includegraphics{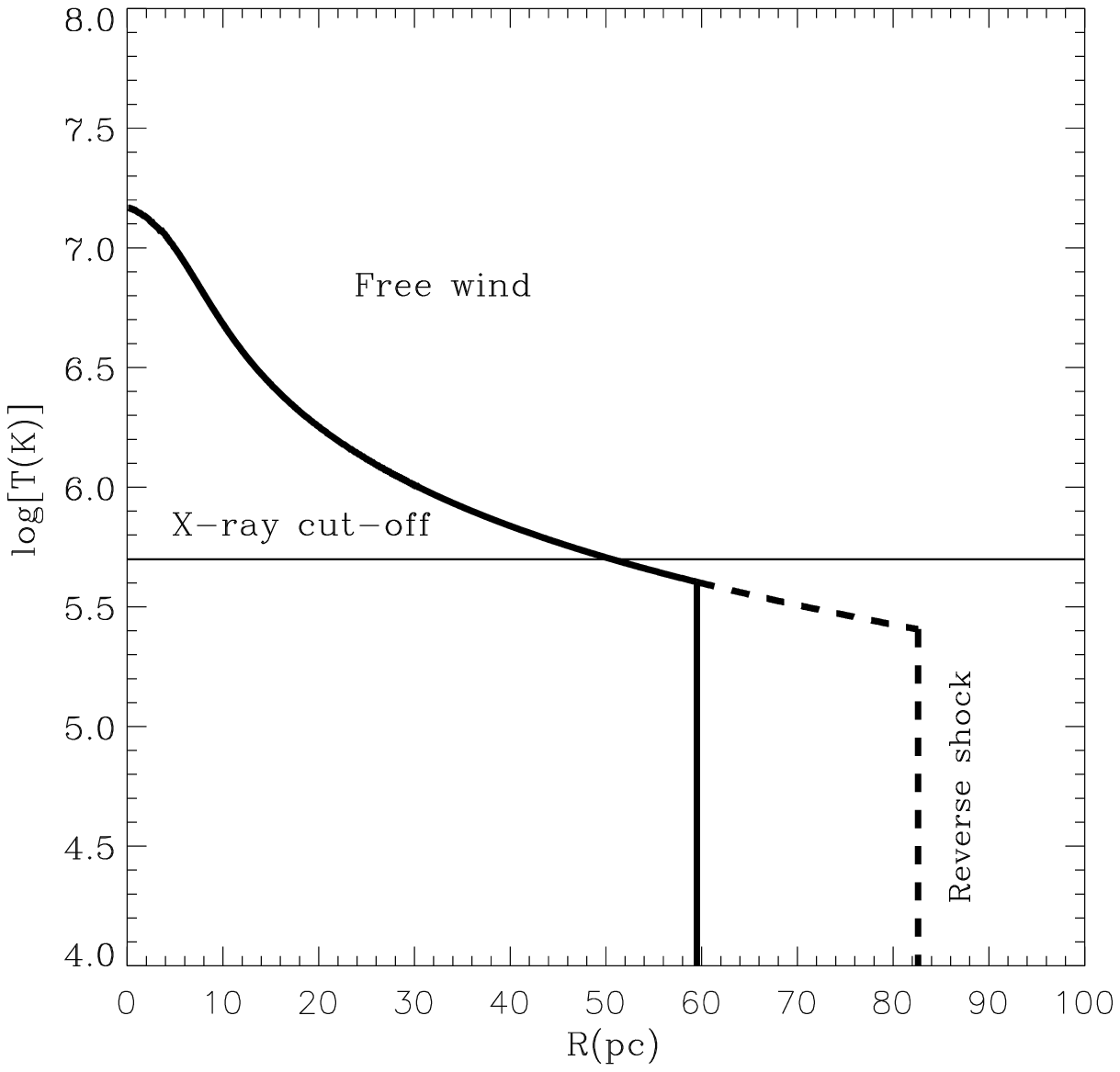}
\includegraphics{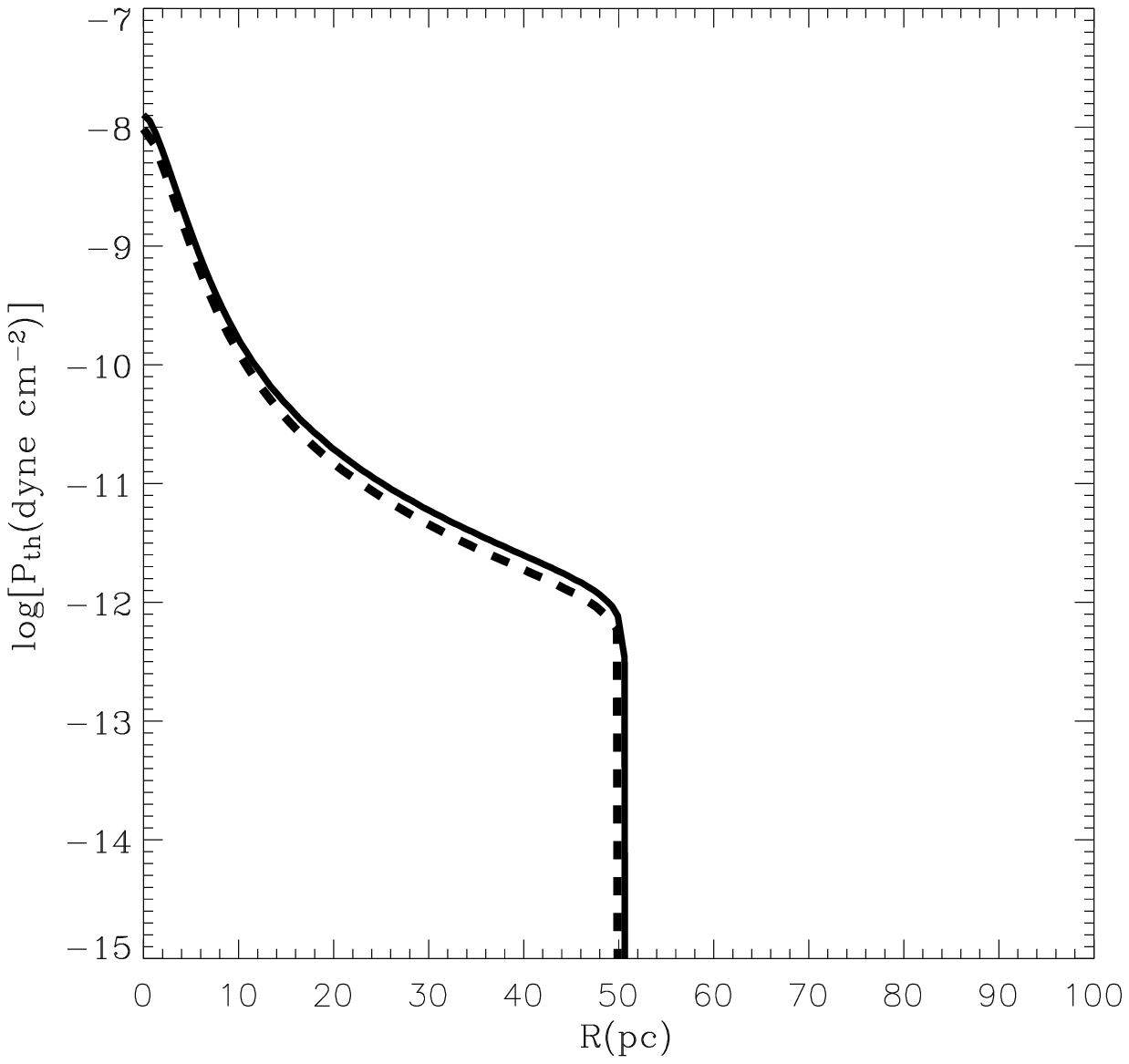}
\caption{The model predicted vs integrated thermal pressure profiles. The
upper and middle panels display the model predicted distributions of thermal 
pressure and temperature, respectively. The thermal pressure profiles obtained
by integration of the model predicted quantities along different lines of 
sight (which one might expect to obtain from the observed X-ray emission) are 
shown in the bottom panels. The left-hand and right-hand side columns 
present the results for the low and high density ($n_{ISM} = 1$~cm$^{-3}$ and 
$n_{ISM} = 1000$~cm$^{-3}$) models, respectively. The solid and dashed lines 
display profiles at different evolutionary times: $t = 2$~Myr and $t = 5$~Myr. 
The horizontal solid lines in the middle panels display the X-ray cut-off 
temperature.}
\label{fig4}
\end{figure}
%---------------------------------------------------------------  
Figure 4 shows that the physical properties of the X-ray plasma derived from 
observations must be taken with care as the distribution of thermal pressure
may be significantly distorted and the true central pressure significantly 
underestimated due to the inhomogeneous distribution of plasma inside the 
wind-driven bubble volume. For example, in the energy dominated regime the 
value of central pressure obtained by integration along lines of sight is 
within a factor of ten smaller than the actual value of the thermal pressure 
at the star cluster center. The integrated profile is significantly shallower 
and does not show a deep gap, which one can notice in the model predicted 
distribution of thermal pressure. The minimum in the integrated distribution 
of thermal pressure is shifted from the model predicted position towards the 
center and therefore it does not mark the position of the reverse shock 
(compare the top and the bottom panels in the left-hand column of Figure 4). 

One can also note in Figure 4 that the shape of the integrated profiles is 
very different in the energy (low density) and momentum (high density) 
dominated cases. In the energy dominated regime the distribution of thermal 
pressure has a strong maximum in the center and a uniform plateau at larger 
radii. In the momentum dominated case the integrated profile is more similar 
to what the free wind model predicts (compare the top and the bottom panels 
in the right-hand side of Figure 4). In this case the major difference 
between the model and the integrated profiles occurs in the center, where  
the model predicted pressure is about ten times larger than that obtained by 
integration, as it also occurs in the energy-dominated case.

Thus, the shape of the thermal pressure profile allows one to distinguish 
between the two possible hydrodynamic regimes. The thermal pressure profile 
with a narrow central spike and a uniform plateau implies that the wind-driven 
bubble evolves in the energy dominated regime and that the reverse shock 
stands within the explored volume. A thermal pressure that drops continuously 
with distance from the star cluster center indicates that the shocked wind 
zone has collapsed due to strong radiative cooling and the wind-driven bubble
evolves in the momentum dominated regime. The reverse shock accelerates then 
out from its energy-dominated position and finally the star cluster wind ends
up impacting directly on the outer shell.

\section{Comparison to other results}

The impact from radiation pressure on the dynamics of giant HII regions has
recently been discussed by \citet{2009ApJ...703.1352K} who claimed that 
radiation pressure might dominate the dynamics of HII regions around massive 
star clusters, whereas we found that radiation has only a narrow window of 
opportunity to dominate the giant HII regions global dynamics.  

The key difference between our results is that \citet{2009ApJ...703.1352K}
did not take into consideration the mechanical power of the exciting cluster
and the time evolution of the star cluster bolometric luminosity. Thus, 
the radiation force in their and our models is compared with different 
pressures: with thermal pressure in the ionized gas behind the ionization 
front in the Krumholz \& Matzner model and with thermal/ram pressure in the
shocked/free wind zone in the wind-driven bubble model. 
Another significant difference is that we consider the time evolution of the 
exciting cluster explicitely that restricts the radiative feedback as the 
bolometric luminosity and the flux of ionizing radiation drop rapidly in 
massive coeval clusters with a normal IMF and ages larger than 3~Myrs.

The impact of radiative heating on the bubble dynamics has been
discussed by \citet{2001PASP..113..677C} and \citet{2012MNRAS.421.1283A}
who have considered bubbles driven by a single massive star. 
As it was stressed by \citet{2001PASP..113..677C}, the real distinction
between the ionizing radiation and the star cluster wind feedback to the
ambient medium is that the characteristic
lifetime of the star is much longer than the wind-blown bubble stalling
time, the time when the bubble expansion velocity drops to the sound speed
value in the ambient medium and the wind-induced shock dissipates. This is,
however, not true in the case when bubbles are driven by massive star clusters
as the characteristic stalling time increases with the mechanical luminosity
of the driving cluster:
%----------------------------------------------------------------------
\begin{eqnarray}
      \label{eq18a}
      & & \hspace{-1.1cm} 
\tau_{stall,ED} = 2.7 c^{-5/2}_{ISM} \left[\frac{(\gamma-1) L_{mech}}
                  {7(9\gamma-4) \pi \rho_{ISM}}\right]^{1/2} \, ,
      \\[0.2cm]     \label{eq18b}
      & & \hspace{-1.1cm}
\tau_{stall,MD} = 0.25 c^{-2}_{ISM} \left[\frac{3 L_{mech}} 
                  {\pi V_{\infty} \rho_{ISM}}\right]^{1/2} \,  ,
\end{eqnarray}
%----------------------------------------------------------------------
where $\tau_{stall,ED}$ and $\tau_{stall,MD}$ are stalling times in the
energy and momentum-dominated regimes, respectively, $c_{ISM}$ is the
sound speed in the ambient medium and we have used an asymptotic ($R_b
>> R_c$) relation for a momentum-dominated bubble expansion velocity
(see equation \ref{eq13b}). In the case of massive star clusters the 
characteristic trapping time $\tau_{trap}$ is 
short compared with both, the wind-driven bubble stalling time, 
$\tau_{stall}$, and the characteristic lifetime of the HII region, 
$\tau_{HII} \approx 10$~Myr (e.g. in our models $\tau_{stall,MD} \approx 
13$~Myr and $\tau_{stall,ED} > 40$~Myr if $T_{ISM} = 1000$K). 
This implies that in this case wind-blown 
bubbles expand for a rather long time in the ``best case for wind'' regime, 
when the wind-driven shell traps all ionizing photons from the central 
source and the leading shock Mach number must be calculated with respect 
to a low sound speed in the neutral ambient medium. 

In the calculations by \citet{2012MNRAS.421.1283A} the ionized shell grows 
thick due to radiative heating what affects the wind-driven bubble dynamics 
because it reduces the volume occupied by the shocked wind gas and thus 
enhances the thermal pressure in the shocked wind zone and behind the leading 
shock front. This effect, however, strongly depends on the mechanical 
luminosity of the driving cluster because the thickness of the ionized shell 
is smaller in systems with a higher internal pressure (compare Figures 2 - 3 
and 4 - 5 in \citealp{2012MNRAS.421.1283A}).
Figure 5 shows the relative thickness of the photoionized 
shell calculated by means of equations (\ref{eq9a}) - (\ref{eq9b}) and 
(\ref{eq12a}). As one can see, in the case of a $10^6$\Msol \, coeval
cluster, the thickness of the ionized shell does not exceed $\approx 10$\% 
of the wind-blown bubble radius what implies that in this case 
radiative heating does not affect significantly the bubble dynamics.
Equations (\ref{eq9a}) - (\ref{eq9b}) and (\ref{eq12a}) predict that
the $\Delta R / R_b$ ratio grows larger for low mass/energetic clusters.
Thus, it is expected that the HII gas filling factor, $f_{HII}$, has to be 
larger in the HII regions developed by the less massive clusters as it is 
the case in the Carina nebula, where $f_{HII} \approx 1$ 
(\citealp{2009ApJ...693.1696H}) and in the 30~Dor region, where the filling 
factor of the ionized gas is very small: $f_{HII} \approx 0.03$ 
(\citealp{1978MNRAS.185..263M}). The effects of 
radiative heating must be also reduced if the wind-driven shells are dusty 
as the dust competes with the gas for ionizing photons what reduces the number
of the incident photons heating the shell.
%---------------------------------------------------------------
\begin{figure}[htbp]
\plotone{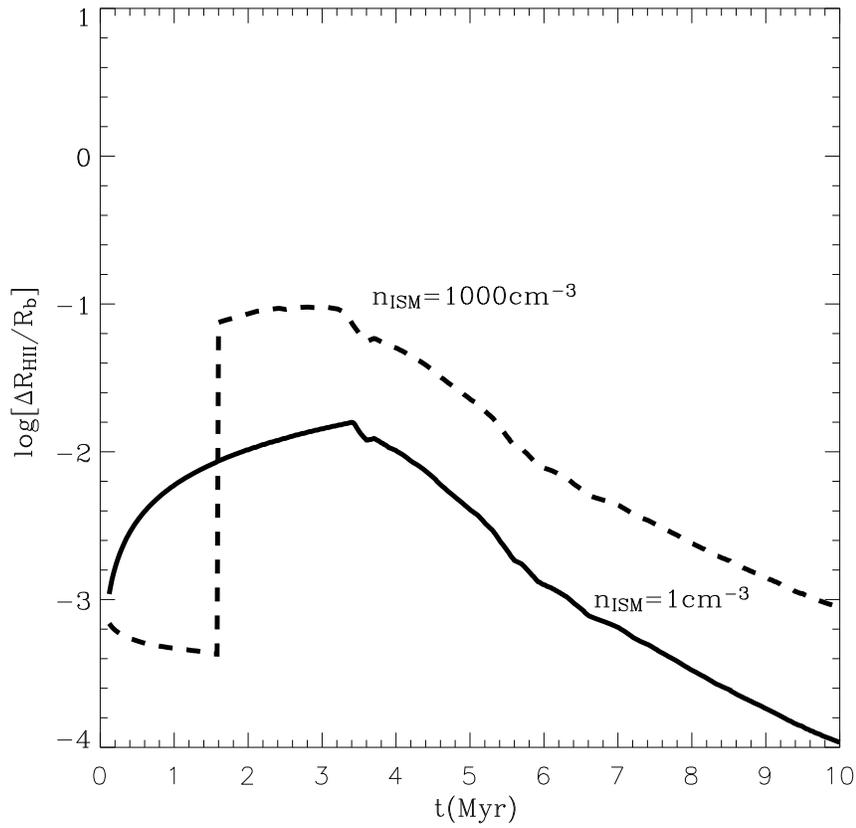}
\caption{The relative thickness of the ionized shell. Solid and dashed lines
present the $\Delta R / R_b$ ratio evolution in the case when the driving 
cluster is embedded into a low (1~cm$^{-3}$) and high (1000~cm$^{-3}$) density
ambient medium, respectively.}
\label{fig5}
\end{figure}
%--------------------------------------------------------------- 

\citet{2009ApJ...693.1696H} discussed the dynamics and the X-ray emission 
from the Carina Nebula bubble and claimed that the leakage of the shocked wind
gas through the holes in the expanding shell may solve a long time standing
problem of the shell growth descipancy what implies that 
\citet{1977ApJ...218..377W} model often overpredicts the sizes and expansion 
velocities of the wind-driven shells (see \citealp{1996ApJ...467..666O}).
The leakage gas model, however, does not work in this 
case as the reverse shock radius calculated by means of equation (\ref{eq5}) 
with the values of thermal pressure, Tr 16 cluster mechanical luminosity and 
mass-loss rate taken from \citet{2009ApJ...693.1696H}: $P_b = 2 \times 
10^{-10}$~dynes cm$^{-2}$, $L_{mech} = 3.5 \times 10^{38}$ erg s$^{-1}$ and 
${\dot M}_{tot} = 1.1 \times 10^{-3}$~\Msol \,  yr$^{-1}$ is comparable or 
even exceeds the observed size of the Carina bubble ($R_{RS} \approx 17$~pc 
while the observed values of the Carina bubble radius range from 10~pc to 
20~pc). 
The leakage bubble model then cannot be used in order to evaluate the 
X-ray luminosity and brightness profile from the Carina Nebula as it 
does not account for the free wind zone (it is assumed that  $R_{RS} << 
R_b$) and thus neglects the contribution of the free wind region to the
X-ray emission. Note also that the apparent discrepancy between the 
\citet{1977ApJ...218..377W} model predictions and the observed bubble radii 
and expansion velocities may be understood if bubbles evolve from flattened 
parental clouds and are observed face-on (see \citealp{1999ApJ...522..863S}).

The geometrical ionization parameter $U = N_{912} / 4 \pi c r^2 n_{HII} = 
N_{912} k \mu_i T_i /  4 \pi c \mu_t P_b$, where $c$
is the speed of light, reaches the maximum value in the momentum-dominated, 
high density model: $log(U_{max}) \approx -1.7$. This value is 
smaller than the maximum value obtained by \citet{2012ApJ...757..108Y},
$log(U_{max}) \approx -1$, who used the \citet{2011ApJ...732..100D}
static model and assumed that the density and thus thermal 
pressure are zero at the inner edge of the strongly illuminated HII region 
(see Appendix B in the paper). This may be the case 
in a static, pressure confined HII regions (\citealp{2011ApJ...732..100D}),
but is not valid  in the case of the wind-driven shell where the thermal 
pressure at the inner edge of the shell must be equal to that in the shocked 
wind zone even if the density distribution in the ionized shell is not 
homogeneous due to strong radiation pressure.

A flat plateau and a steep central spike predicted by the energy-dominated
wind-driven bubble model agree surprisingly well with the shape of the
thermal pressure profile of 30~Dor region obtained by 
\citet{2011ApJ...731...91L}. This allows one to rule out the momentum 
dominated regime and conclude that the leakage of the hot gas from the 30~Dor 
shell is not a significant factor as suggested in \citet{2011ApJ...731...91L}.
This conclusion also agrees with the detailed analysis of the X-ray emission 
from the 30~Dor region by \citet{1991ApJ...370..541W} and 
\citet{2006AJ....131.2140T} who claimed that the diffuse X-ray 
emission is enveloped and probably confined by the cooler gas which outlines 
the classic picture of 30~Dor and extends for about 300~pc from north to 
south. 

Radiation pressure is approximately equal to the gas pressure in the
highly ionized ridge in the center of 30~Dor region. Outside of this 
central zone the $P_{rad}/P_{gas}$ ratio drops below 1/3 
(\citealp{2011ApJ...738...34P}) and the ionization parameter $U$ does not show 
much gradient (\citealp{2009ApJ...694...84I}). The $P_{rad}/P_{gas} < 1/3$ 
value is roughly consistent with the wind-driven model predictions (see 
Figure 3), if one bears in mind the complicated history of star formation in 
the 30~Dor starburst with three episodes of star formation of increasing 
intensity which took place approximately 5~Myr, 2.5~Myr and about 1.5~Myr ago
(\citealp{1999A&A...347..532S}). The radius of the central arc ($\sim 10$~pc) is
comparable to the size of NGC 2070, the major driving cluster in the 30~Dor 
starburst, what requires a more realistic than a spherical shell illuminated 
by a central source model as was also stressed by \citet{2007ApJ...669..269S}
when discussing the large values of the ionization parameter 
($log(U) \ge -1.53$) detected in the HII regions around two young (ages 
less than 3~Myr) stellar clusters in the Antenna.  

Our calculations thus favor the results by \citet{2011ApJ...738...34P} who 
obtained a much lower radiation over thermal pressure ratio in the 30~Dor
region than \citet{2011ApJ...731...91L}. Note also that Pellegrini et al. 
results look in better agreement with the spectroscopic virial mass of the 
30~Dor cluster obtained by \citet{2009AJ....137.3437B} ($M_{vir} \approx 9 
\times 10^5$\Msol) which leads, in the case of a coeval cluster with a Kroupa 
IMF and $0.4Z_{\odot}$ metallicity, to the maximum value of bolometric 
luminosity $4.4 \times 10^{42}$~erg s$^{-1}$ what is about two times smaller 
than the $L_{bol}$ obtained by \citet{2011ApJ...731...91L}.

\section{Conclusions}

Here we have explored the contribution of radiative and dynamical (thermal or 
ram) pressure to the dynamics of giant HII regions and found only a narrow 
window of opportunity for radiation pressure to be a dominant factor.
We stress the importance of the mechanical feedback from the exciting
cluster and the radiation power time evolution, and derive analytic relations 
which show how the radiation over dynamical pressure ratio evolves with
time. Careful analysis of the two extreme models which cover conditions 
ranging from the interstellar medium in normal galaxies to those found 
in dense giant molecular clouds, led us to conclude that the dynamical 
pressure dominates always after about 3~Myr, if one uses the standard
bubble and star cluster synthesis models. By that time the major factor 
which defines the relative contribution from the two driving forces, the 
bolometric over the mechanical luminosity ratio, drops within a factor of ten 
(see Figure 1) and the impact of radiation pressure becomes soon negligible 
even in the most preferable case. We want to stress that the reduction
of thermal pressure in the shocked wind zone due to leakage of the hot 
shocked plasma out of a porous shell cannot change this conclusion as may 
only lead to the displacement of the reverse shock from the 
\citet{1977ApJ...218..377W} model predicted position, but does not affect the 
distribution of the ram pressure in the free wind zone. 
In the extreme case this may lead to the momentum dominated expansion as it is 
the case if the shocked wind zone collapses due to strong radiative cooling
or if the reverse shock is not formed due to the large proton mean free path 
in the shocked wind region (\citealp{2001PASP..113..677C}). This leads to the 
maximum value of the $P_{rad}$ over $P_b$ ratio which is shown in Figure 3 
and occurs when the wind-driven bubble evolves in the momentum-dominated 
regime. However, even in this ``best case for radiation'' regime the 
contribution from radiation pressure becomes small after about 3~Myr (see 
dashed line in Figure 3). We thus conclude that radiation pressure, 
despite being significant during the earlier evolution, in general has a poor 
impact on the expansion of giant shells powered by massive star clusters 
unless the Starburst99 synthesis model overestimates the star cluster 
mechanical luminosity significantly or a significant fraction of the 
mechanical energy is lost inside the star cluster volume where the 
stellar mechanical energy is thermalized in random collisions of nearby 
stellar winds and supernovae ejecta.

\acknowledgments 
We thank our anonymous referee for a detailed report and many valuable
suggestions which greatly improved the paper. 
This study has been supported by CONACYT - M\'exico, research grants
131913 and 167169 and by the Spanish Ministry of Science and Innovation  
under the collaboration ESTALLIDOS (grant AYA2007-67965-C03-01) and 
Consolider-Ingenio 2010 Program grant CSD2006-00070: First Science 
with the GTC. 

\bibliographystyle{aa}
\bibliography{30Dor}

\end{document}